\documentstyle[eqsecnum,aps,epsf]{revtex}      
\newcommand{\postscript}[2] {\setlength{\epsfxsize}{#2\hsize}

\centerline{\epsfbox{#1}}}

\begin{document}
\twocolumn[\hsize\textwidth\columnwidth\hsize\csname @twocolumnfalse\endcsname

\title{\bf Phase transition from a $d_{x^2-y^2}$ to 
$d_{x^2-y^2}+d_{xy}$  superconductor}

\author{Angsula Ghosh and Sadhan K Adhikari}
\address{Instituto de F\'{\i}sica Te\'orica, 
Universidade Estadual Paulista,
01.405-900 S\~ao Paulo, S\~ao Paulo, Brazil}

\date{\today}
\maketitle

\begin{abstract}

We study  the phase transition from a $d_{x^2-y^2}$ to
$d_{x^2-y^2}+d_{xy}$ superconductor using the tight-binding model of
two-dimensional cuprates. As the temperature is lowered past the critical
temperature $T_c$, first a $ d_{x^2-y^2}$ superconducting phase is
created. With further reduction of temperature, the $ d_{x^2-y^2}+d_{xy}$
phase is created at temperature $T=T_{c1}$. We study the temperature
dependencies of the order parameter, specific heat and spin susceptibility
in these mixed-angular-momentum states  on square lattice and on a lattice
with orthorhombic distortion. 
The above-mentioned phase transitions are identified by two jumps in
specific heat at $T_c$ and $T_{c1}$.

{PACS number(s): 74.20.Fg, 74.62.-c, 74.25.Bt}

\end{abstract} 

\vskip1.5pc]


Inspite of many theoretical and experimental studies on high-$T_c$
cuprates
the exact symmetry of
the order parameter  is still a subject of active research  \cite{n1}.
 However, there is  evidence that
the cuprates have
singlet $d$-wave Cooper pairs and the order parameter has $d_{x^2-y^2}$
symmetry in two dimensions \cite{n1}.  Recent measurements \cite{h} of 
penetration depth and superconducting specific heat at different
temperatures $T$ and related theoretical analyses \cite{t1,c} also support
this.  However, several phase-sensitive measurements of the
order parameter of the cuprates indicate a significant mixing of a
distinct angular momentum component with a predominant $d_{x^2-y^2}$ state
at  temperatures below a second critical temperature $T_{c1}$. For
temperatures between $T_{c1}$ and $T_c$ only the $d_{x^2-y^2}$ state
survives.  Below $T_{c1}$ the order parameter can have a mixed-symmetry
state of type $d_{x^2-y^2}+\exp(i\theta)\chi$,  where $\chi$ represents a
state of different symmetry. The most probable possibilities for $\chi$
are the $s$ or  $d_{xy}$ wave.  The possibility of a mixed $(s-d)$-wave
symmetry was first suggested theoretically by Ruckenstein {\it et al.} and
Kotliar \cite{6}.

There are experimental evidences based on Josephson supercurrent for
tunneling between a conventional $s$-wave superconductor (Pb) and twinned
or untwinned single crystals of YBa$_2$Cu$_3$O$_7$ (YBCO)  that YBCO has
mixed $d_{x^2-y^2}\pm s$ or $d_{x^2-y^2}\pm is$ symmetry \cite{5} at lower
temperatures. Recently, the existence of these mixed-symmetry states has
been explored to explain the nuclear magnetic resonance data in the
superconductor YBCO and the Josephson critical current observed in
YBCO-SNS and YBCO-Pb junctions \cite{8}.

 Kouznetsov {\it et al.} \cite{K} performed some c-axis Josephson
tunneling experiments by depositing conventional superconductor (Pb) 
across a single twin boundary of a YBCO crystal. By measuring the critical
current as a function of the angle and magnitude of a magnetic field
applied in the plane of the junction they also found the evidence of a
mixed-symmetry order parameter in YBCO involving $d_{x^2-y^2}$ and $s$
waves. By measuring the microwave complex conductivity in the
superconducting state of high quality YBa$_2$Cu$_3$O$_{7-\delta}$ single
crystals at 10 GHz using a high-Q Nb cavity Sridhar $et al$ also suggested
the existence of a multicomponent superconducting order parameter in YBCO
\cite{S}.

A similar conclusion of the existence of mixed-symmetry states 
may also be obtained based on the results of angle-resolved
photoemission spectroscopy experiment by Ma {\it et al.} in which a
temperature dependent gap anisotropy in oxygen-annealed
Bi$_2$Sr$_2$CaCu$_2$O$_{8+x}$ was detected \cite{7}. The measured gaps
along directions $\Gamma-M$ and $\Gamma-X$ are non-zero at low
temperatures and their ratio was strongly temperature dependent. Using
Ginzburg-Landau theory, Betouras and Joynt \cite{bj1} demonstrated that
one way of explaining this behavior is to employ a mixed-symmetry state of
the $d_{x^2-y^2}+s$-wave type. They also conclude that the actual symmetry
of the order parameter should vary substantially from one compound to
another and for different levels of doping. This also suggests the
possible appearance of a $d_{x^2-y^2}+d_{xy}$ state under favorable
conditions.

More recently, Krishana {\it et al.} \cite{Kr} reported a phase transition
in the  superconductor Bi$_2$Sr$_2$CaCu$_2$O$_8$ induced by a
magnetic field from a study of the thermal conductivity as a function of
temperature and applied field.
Laughlin \cite{L} provided a theoretical explanation of the
observation by Krishana {\it et al.} \cite{K} that 
for weak magnetic field a time-reversal symmetry breaking state of mixed
symmetry is induced in Bi$_2$Sr$_2$CaCu$_2$O$_8$. From a study of vortex
in a $d$-wave superconductor using a self-consistent Bogoliubov-de Gennes
formalism, Franz and Te\'sanovi\'c \cite{L} also predicted the possibility
 of a superconducting state of mixed symmetry. 
This mixed-symmetry state is likely to be 
 a minor $s$ or $d_{xy}$ component superposed  
on a $d_{x^2-y^2}$ state for $T< T_{c1}$. 
 
From different experimental observations it is now generally accepted that
a time-reversal symmetry breaking state of type $d_{x^2-y^2}+i \chi$
is possible in the presence of an
external field or magnetic impurity. This mixed-symmetry state is observed
close to these impurities, surfaces/twin boundaries in the ab-plane or
vortices. The nature of the mixed state varies from compound to compound. 
There are physical reasons for the appearance of these states. Either
spin-orbit coupling with magnetic impurities or Andreev reflected bound
states which create internal currents at the boundaries is responsible for
these states \cite{az}. However, orthorhombicity plays a crucial role in
the generation of time-reversal symmetric mixed states. For example, it is
established from a Ginzburg-Landau functional analysis \cite{bj2} that the
orthorhombicity has a consequence in the development of a $d + s$ state
instead of a time-reversal symmetry broken one. Moreover, from a
theoretical point of view, time-reversal symmetric states of type
$d_{x^2-y^2}+\chi$  are expected to
be
allowed depending on the orthorhombic distortion.

There have been some  studies \cite{9a} on the phase
transition to a $d_{x^2-y^2}+\exp(i\theta)\chi$ phase from a $d_{x^2-y^2}$
phase with $\theta = \pi /2$ and $\chi = d_{xy}$ or a $s$ state. From
theoretical considerations we find that there are two possibilities for
the phase $\theta$:  0 or $\pi/2$.  For $\theta = 0$, we find numerically
that there is no stable $d_{x^2-y^2}+s$ phase. Here we study the phase
transition to a $d_{x^2-y^2}+d_{xy}$ phase from a 
$d_{x^2-y^2}$ phase below  $T_{c1}$. In particular we study 
 the temperature dependencies of the order
parameter, specific heat, and spin susceptibility in the mixed-symmetry
state.

There is no suitable microscopic theory for high-$T_c$ superconductors and
there is controversy about a proper description of the normal state and
the pairing mechanism for such materials \cite{n1}.  In the absence of a
 microscopic theory, a phenomenological tight-binding model in two
dimensions with the proper lattice symmetry will be used   \cite{P}. This
model has been  successful
in describing many properties of high-$T_c$ materials.

We study the temperature dependencies of specific heat and
susceptibility of a $ d_{x^2-y^2}+d_{xy}$-wave superconductor with a
weaker $d_{xy}$ wave both on square lattice and on a lattice with
orthorhombic distortion. The order parameter of a
$d_{x^2-y^2}+d_{xy}$-wave superconductor has nodes on the Fermi surface
and changes sign across it, and consequently, its superconducting
observables also exhibit power-law dependencies on temperature.  On the
other hand, the order parameters for the mixed $ d_{x^2-y^2}+is$ and
$d_{x^2-y^2}+id_{xy}$-wave states do not have a node on the Fermi surface
and the corresponding observables have a exponential dependencies on
temperature. In the present study on $d_{x^2-y^2}+d_{xy}$-wave states the
specific heat exhibits two jumps at $T=T_{c1}$ and $T=T_c$, which clearly
exhibits the phase transition at $T_{c1}$.

In  the 
 present two-dimensional tight binding model
the effective interaction $V_{{\bf k}{\bf
q}}$ for transition from a momentum ${\bf q}$ to ${\bf k}$ is taken to be
separable, and is expanded in terms of some general basis functions
$\eta_{i{\bf k}}$, labelled by the index $i$, as $V_{{\bf k}{\bf q}}=
-\sum_i V_i \eta_{i{\bf k}}\eta_{i{\bf q}}$ \cite{fn}. The functions
$\eta_{i{\bf k}}$ are associated with a one dimensional irreducible
representation of the point group of square lattice $C_{4v}$ and are
appropriate generalizations of the circular harmonics incorporating the
proper lattice symmetry.
The effective interaction after
including the two appropriate basis functions for singlet pairing is taken
as \begin{equation}\label{2} V_{{\bf k}{\bf q}}=-V_1\eta_{1{\bf k}}
\eta_{1{\bf q}}- V_2 \eta_{2{\bf k}}\eta_{2{\bf q}}, \end{equation} where
$\eta _{1\bf q} \equiv (\cos q_x -\beta \cos q_y)$ corresponds to
$d_{x^2-y^2}$ symmetry, $\eta _{2\bf q}\equiv \sin q_x \sin q_y$
corresponds to $d_{xy}$ symmetry, and where $\beta =1$ corresponds to a
square lattice, and $\beta \ne 1$ represents orthorhombic distortion.  In
this case the quasiparticle dispersion relation is given by
$ \epsilon_{\bf k}=-2t[\cos k_x+\beta \cos
k_y-\gamma\cos k_x \cos k_y],$  where $t$ and $\beta t$ are
the nearest-neighbour hopping integrals along the in-plane $a$ and $b$
axes, respectively, and $\gamma t/2$ is the second-nearest-neighbour
hopping integral. The energy $\epsilon_{\bf k}$ is measured with respect
to  the Fermi   surface.

At a finite $T$, one has the following BCS
equation \begin{eqnarray} \Delta_{\bf k}& =& -\sum_{\bf q} V_{\bf
kq}\frac{\Delta_{ \bf q}}{2E_{\bf q}}\tanh \frac{E_{\bf q} }{2k_BT}
\label{130} \end{eqnarray} with $E_{\bf q} = [(\epsilon_{\bf q} - E_F )^2
+ |\Delta_{\bf q}|^{2}]^ {1/2},$ where $E_F$ is the Fermi temperature and
$k_B$ the Boltzmann's constant. The order parameter 
has the following  anisotropic form: \begin{equation} \Delta _{\bf
q} \equiv \Delta_1 \eta _{1\bf q} +C\Delta_2 \eta _{2\bf q},
\label{op}\end{equation} where $C$ is a complex number of unit modulas
$|C|^ 2=1$. If we substitute Eqs. (\ref{2}) and (\ref{op}) into the BCS
equation (\ref{130}), one can separate the resultant equation in its real
and imaginary parts. The resultant equations only have solution for real
$\Delta_1$ and $\Delta_2$, when the complex parameter $C$ is either purely
real or purely imaginary.  The solution for purely imaginary $C$, e.g.,
$C=i$ have been extensively studied in relation to mixed $d_{x^ 2-y^
2}+is$ and $d_{x^ 2-y^ 2}+id_{xy}$ states \cite{9a}. Here we consider
the solution for $C=1$, for $d_{x^ 2-y^ 2}+d_{xy}$ state. Using the form
(\ref{op}) of $\Delta_{\bf q}$ with $C=1$ and potential (\ref{2}), Eq. 
(\ref{130}) becomes the following coupled set of BCS equations
\begin{eqnarray} \Delta_1=V_1\sum_{\bf q}\frac{\eta _{1\bf q}[\Delta_1\eta
_{1\bf q}+\Delta_2 \eta _{2\bf q}]}{2E_{\bf q}}\tanh \frac{E_{\bf
q}}{2k_BT}\label{131} \end{eqnarray} \begin{eqnarray}
\Delta_2=V_2\sum_{\bf q}\frac{\eta _{2\bf q}[\Delta_1\eta _{1\bf q}
+\Delta_2 \eta _{2\bf q}]}{2E_{\bf q}}\tanh \frac{E_{\bf
q}}{2k_BT},\label{132} \end{eqnarray} where both the interactions $V_1$
and $V_2$ are assumed to be energy-independent constants for
$|\epsilon_{\bf q} - E_F| < k_B T_D$ and zero for $|\epsilon_{\bf q} -
E_F| > k_B T_D$, where $k_B T_D$ is a 
 purely mathematical cutoff  introduced to eliminate the
ultraviolet divergence in the BCS equation and should be compared with the
physically motivated Debye cutoff in the case of the conventional
superconductors.

We solved the coupled set of equations (\ref{131}) and (\ref{132}) 
numerically and calculated the gaps $\Delta_1$ and $\Delta_2$ at various
temperatures for $T<T_c$. We have performed calculations (1) on a perfect
square lattice and (2)  in the presence of an orthorhombic distortion with
cut off $k_BT_D=0.02586 $ eV ($T_D = 300$ K) in both cases. The parameters
for these two cases are the following:  (1) Square lattice $-$ (a) 
$t=0.2586 $ eV, $\beta=1$, $\gamma = 0$, $V_{2}=8.5t$, and $V_1=0.73t$,
$T_c = 71$ K, $T_{c1}$ = 28 K; (b) $t=0.2586 $ eV, $\beta=1$, $\gamma =
0$, $V_{2}=9.0t$, and $V_1=0.73t$, $T_c = 71$ K, $T_{c1}$ = 55 K; (2) 
Orthorombic distortion $-$ (a)  $t=0.2586 $ eV, $\beta = 0.95$, and
$\gamma=0$, $V_{2}=8.35t$, and $V_1=0.97t$, $T_c$ = 70 K, $T_{c1} $ = 30
K;  (b)  $t=0.2586 $ eV, $\beta = 0.95$, and $\gamma =0$, $V_{2}=8.7t$,
and $V_1=0.97t$, $T_c$ = 70 K, $T_{c1} $ = 50 K.  For a very weak $d_{x^
2-y^ 2}$-wave ($d_{xy}$-wave) coupling the only possible solution
corresponds to $\Delta_1 =0$ ($\Delta_2 =0$).  We have studied the
solution only when a coupling is allowed between Eqs. (\ref{131}) and
(\ref{132}).

\vskip -2.8cm
\postscript{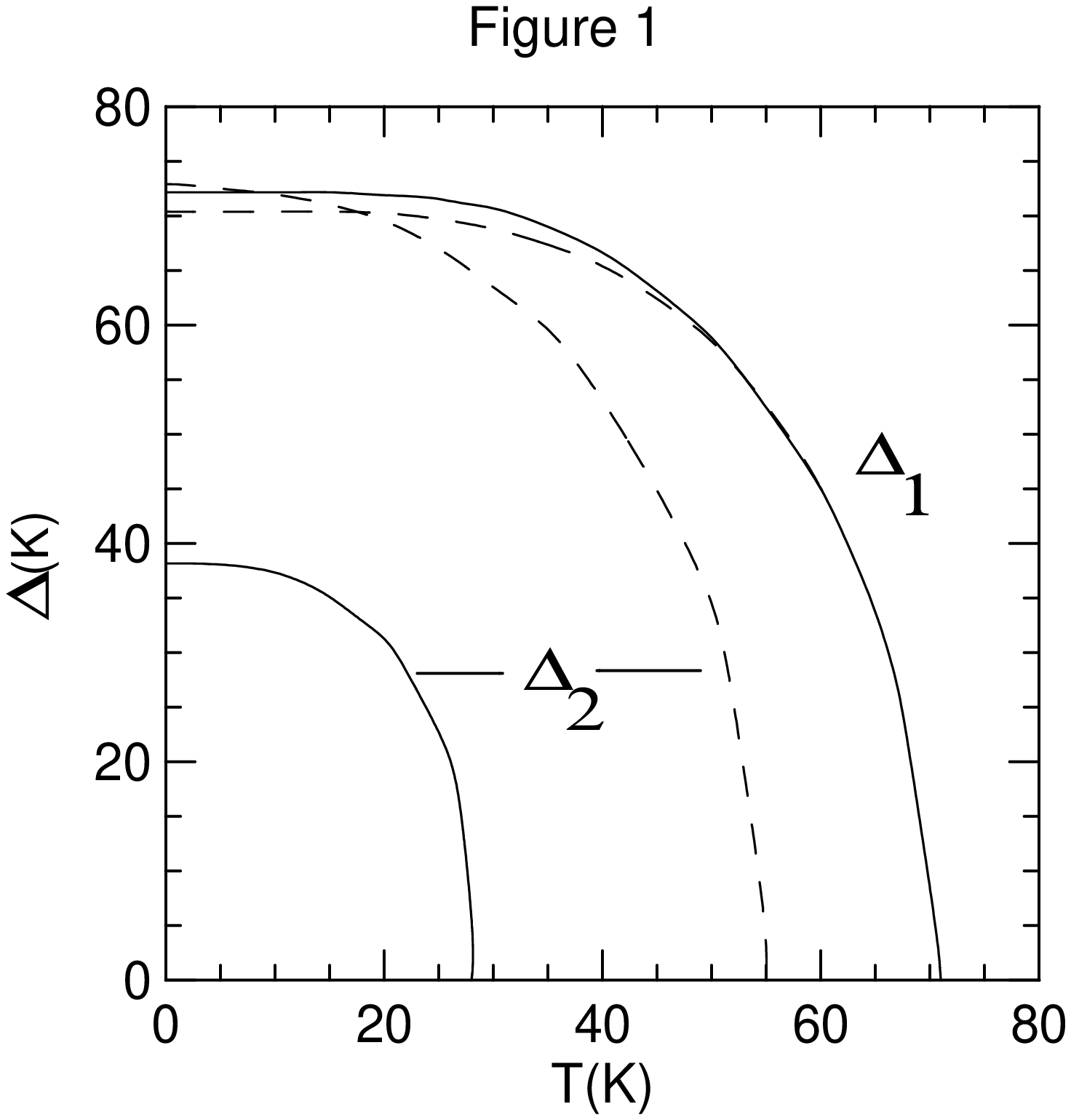}{1.0}    
\vskip -2.1cm

{ {\bf Fig. 1.}   The  order parameters
$\Delta_1$,  $\Delta_2$ in Kelvin 
 at different temperatures for  models 
1(a) (full line) and 1(b) (dashed line) on square lattice  
described in text.}

\vskip -2.8cm
\postscript{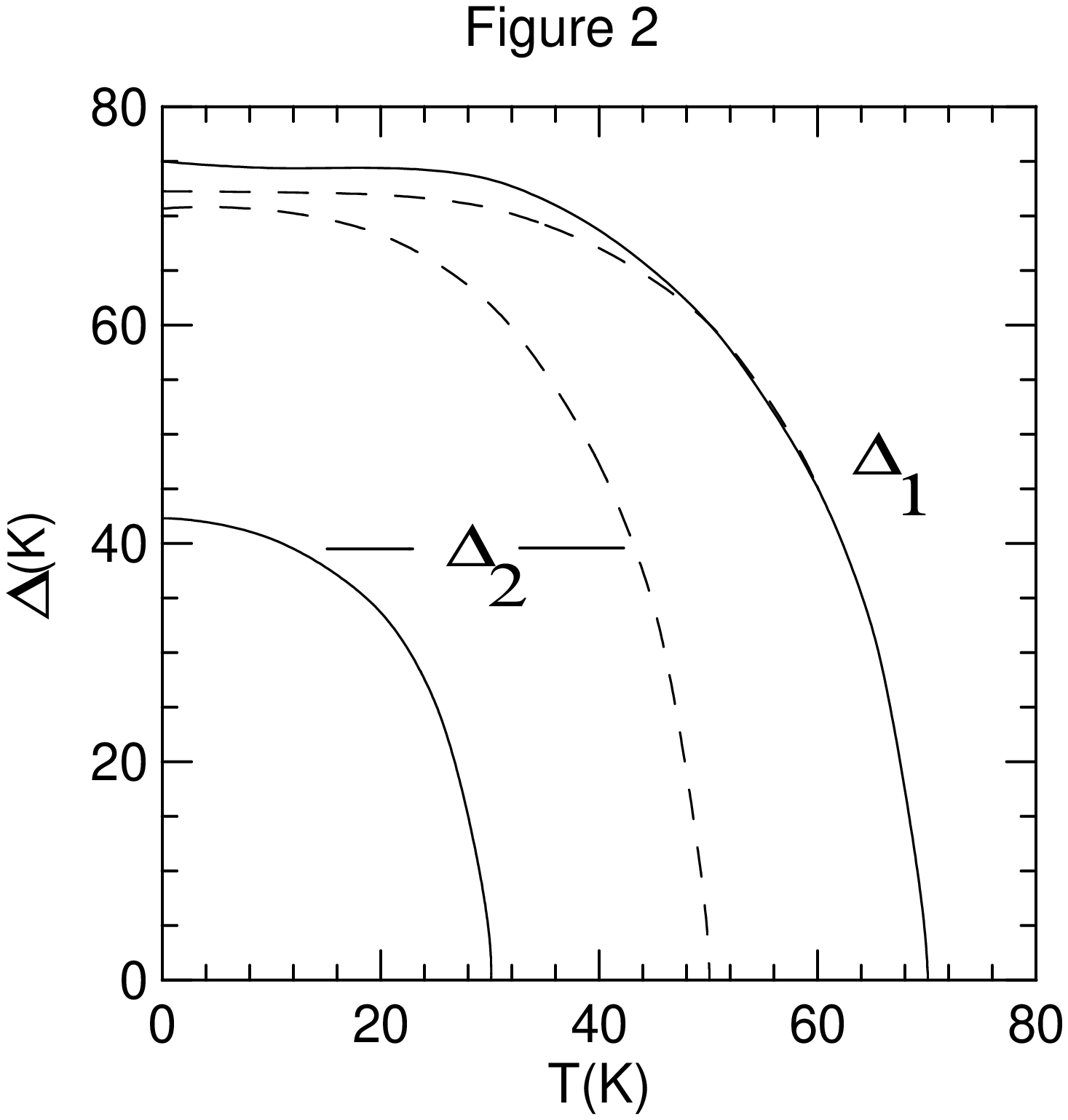}{1.0}    
\vskip -2.1cm

{ {\bf Fig. 2.}   The  same as in Fig. 1 
 at different temperatures 
for  models 
2(a) (full line) and 2(b) (dashed line) in the presence of orthorhombic
distortion   
described in  text.}

\vskip .3cm

In Figs.  1 and 2 we plot the temperature dependencies of different
$\Delta$'s for the following two sets of $d_{x^2-y^2}+d_{xy}$ wave
corresponding to models 1 and 2, respectively.  In all cases, with
the lowering of temperature passed $T_c$, the parameter $\Delta_1$
increases up to $T=T_{c1}$. As $T$ is lowered further, $\Delta_2$ becomes
nonzero at $T=T_{c1} $ and begins to increase. As temperature is lowered,
both $\Delta_1$ and $\Delta_2$ first increase and then attain a constant
value at zero temperature.

The different superconducting and normal specific heats are plotted in
Figs. 3 and 4 for square lattice [models 1(a) and 1(b)] and orthorhombic
distortion [models 2(a) and 2(b)], respectively.  In both cases the
specific heat exhibits two jumps $-$ one at $T_c$ and another at $T_{c1}$. 
From  Figs.  1 and 2 we see
that the temperature

\vskip -3.0cm
\postscript{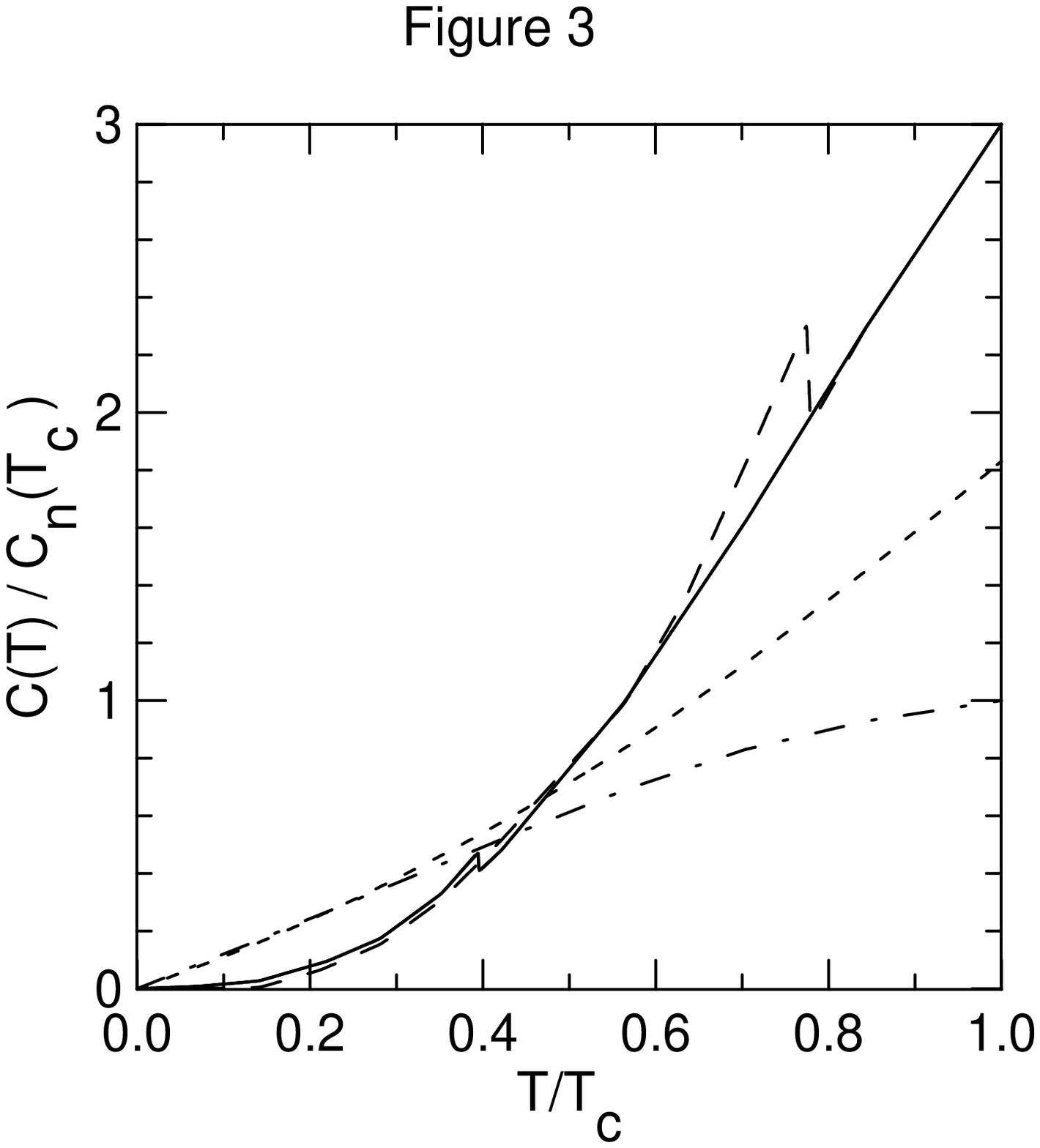}{1.0}    
\vskip -1.5cm

{ {\bf Fig. 3.}  Specific heat ratio $C(T)/C_n(T_c)$ versus $T/T_c$  for
models 1(a)
and 1(b) on square lattice: 1(a)
(full line), 1(b) 
 (dashed line), $d_{xy}$ (dotted line), 
normal (dashed-dotted line). }

\vskip -3.0cm
\postscript{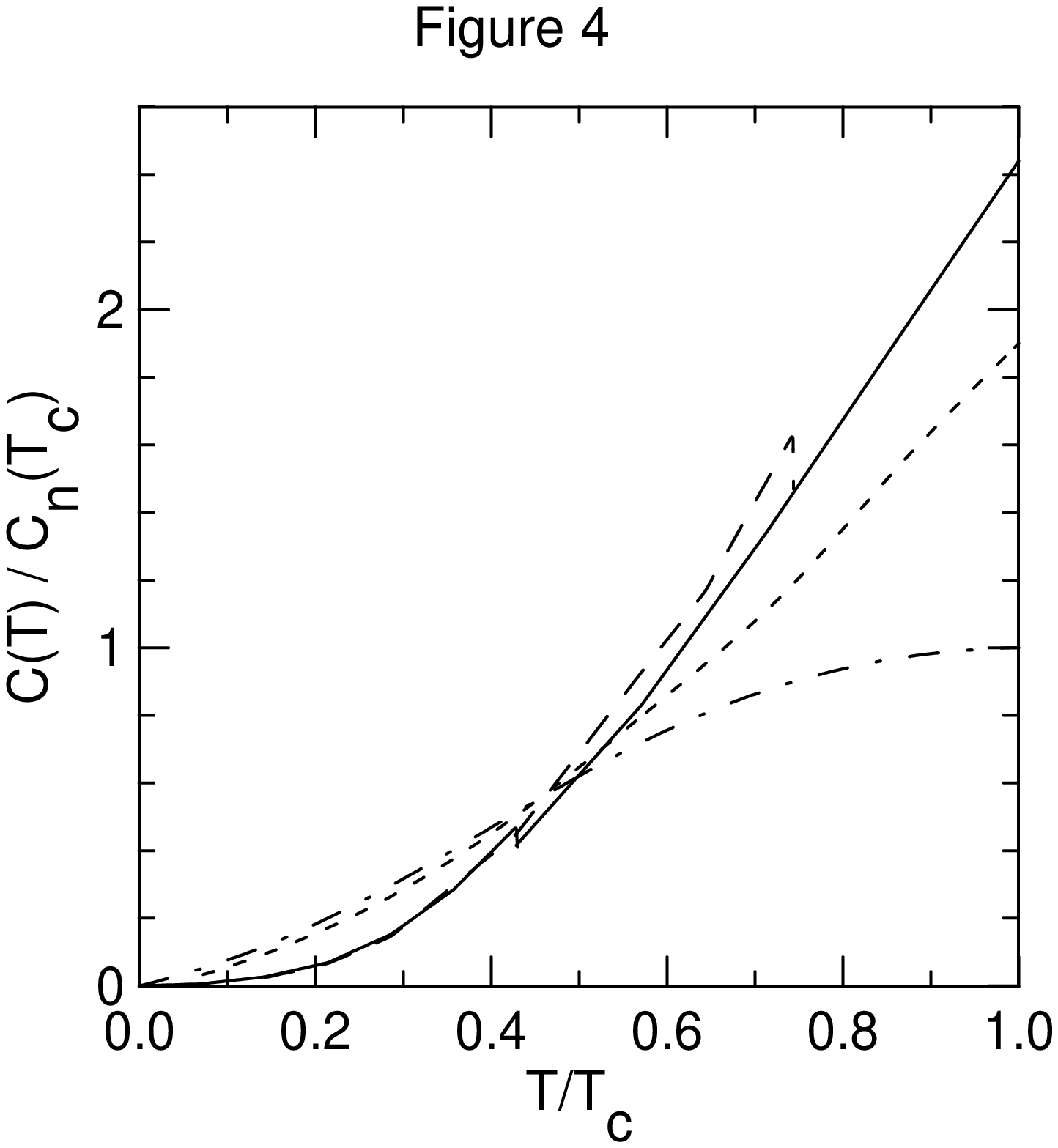}{1.0}    
\vskip -1.5cm

{ {\bf Fig. 4.}  Specific heat ratio $C(T)/C_n(T_c)$ versus $T/T_c$  for
models 2(a)
and 2(b) in the presence of orthorhombic distortion: 2(a)
(full line), 2(b) 
 (dashed line), $d_{xy}$ (dotted line), 
normal (dashed-dotted line).}

\vskip .3cm

\noindent derivative of $|\Delta_{\bf q}|^2$ has discontinuities at $T_c$
and
$T_{c1}$ due to the vanishing of $\Delta_1$ and $\Delta_2$, respectively,
responsible for the two jumps in specific heat (see, definition in 
Ref. \cite{c}).  For
a pure $d_{x^2-y^2}$
wave we find that the specific heat exhibits a power-law dependence on
temperature. However, the exponent of this dependence varies with
temperature. For small $T$ the exponent is approximately 2.5, and for
large $T$ ($T\to T_c$) it is nearly 2. For the mixed
$d_{x^2-y^2}+d_{xy}$-wave model, for $T_c > T > T_{c1}$ the specific heat
exhibits $d$-wave power-law behavior. For $d$-wave models
$C_s(T_c)/C_n(T_c)$ is a function of $T_c$ and $\beta$. In Figs. 2 and 3
this ratio for the $d_{x^ 2-y^ 2}$-wave case, for $T_c$ = 70 K, is
approximately 3 (2.44) for $\beta =$ 1 (0.95).  For the $d_{xy}$-wave
case, for $T_c$ = 70 K, this ratio is approximately 1.81 (1.9) for $\beta
=$ 1 (0.95).  In a continuum calculation this ratio was 2 in the absence
of a van Hove singularity \cite{c}.

\vskip -3.0cm
\postscript{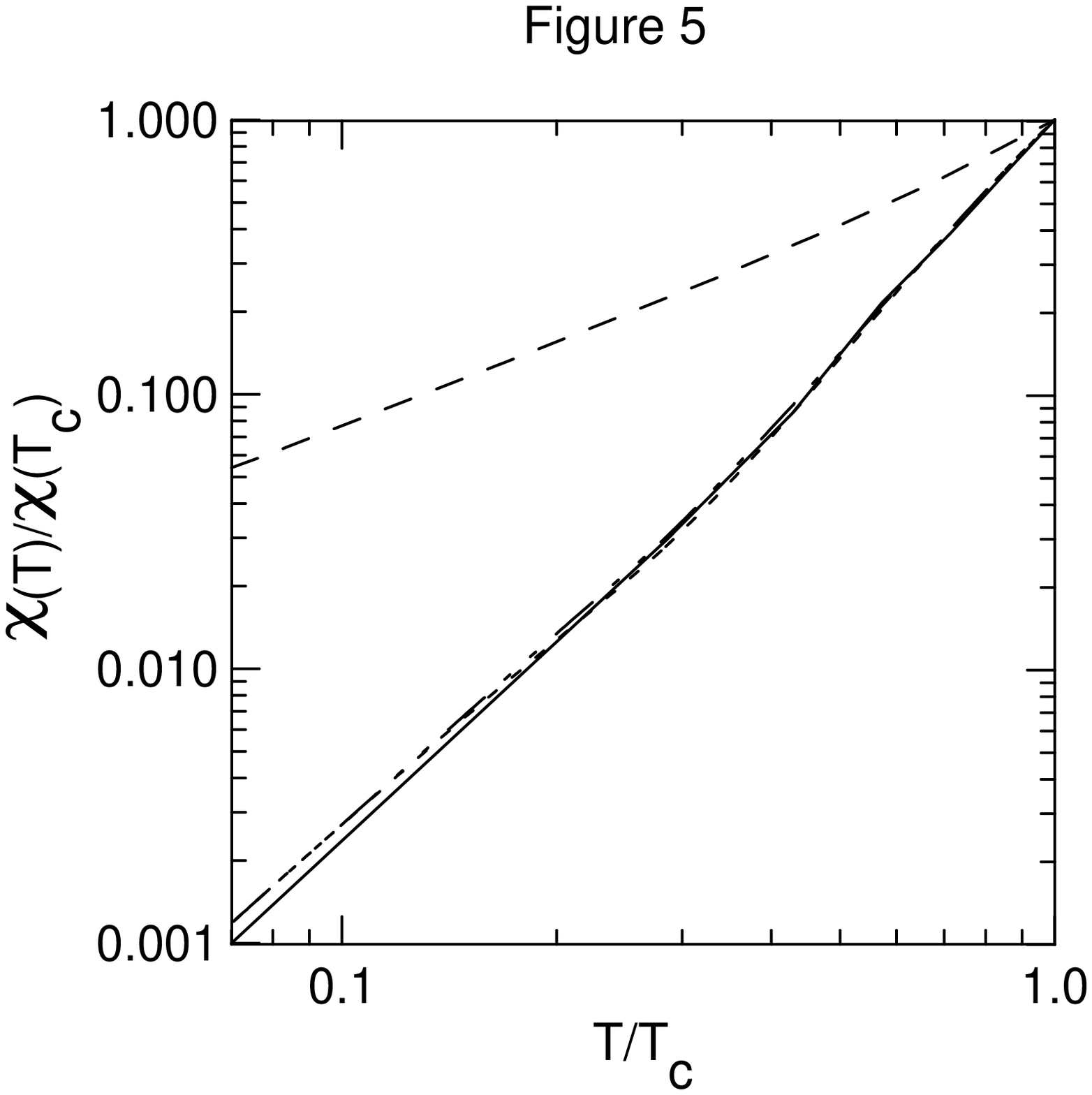}{1.0}    
\vskip -1.5cm

{{\bf Fig. 5.}  Susceptibility  ratio $\chi(T)/\chi(T_c)$ versus $T/T_c$  
for square lattice: 
pure $d_{x^2-y^2}$ wave (solid line),
pure $d_{xy}$ wave (dashed line), model 1(a) (dotted line), and 
model 1(b) (dashed-double-dotted line).}

\vskip -3.0cm
\postscript{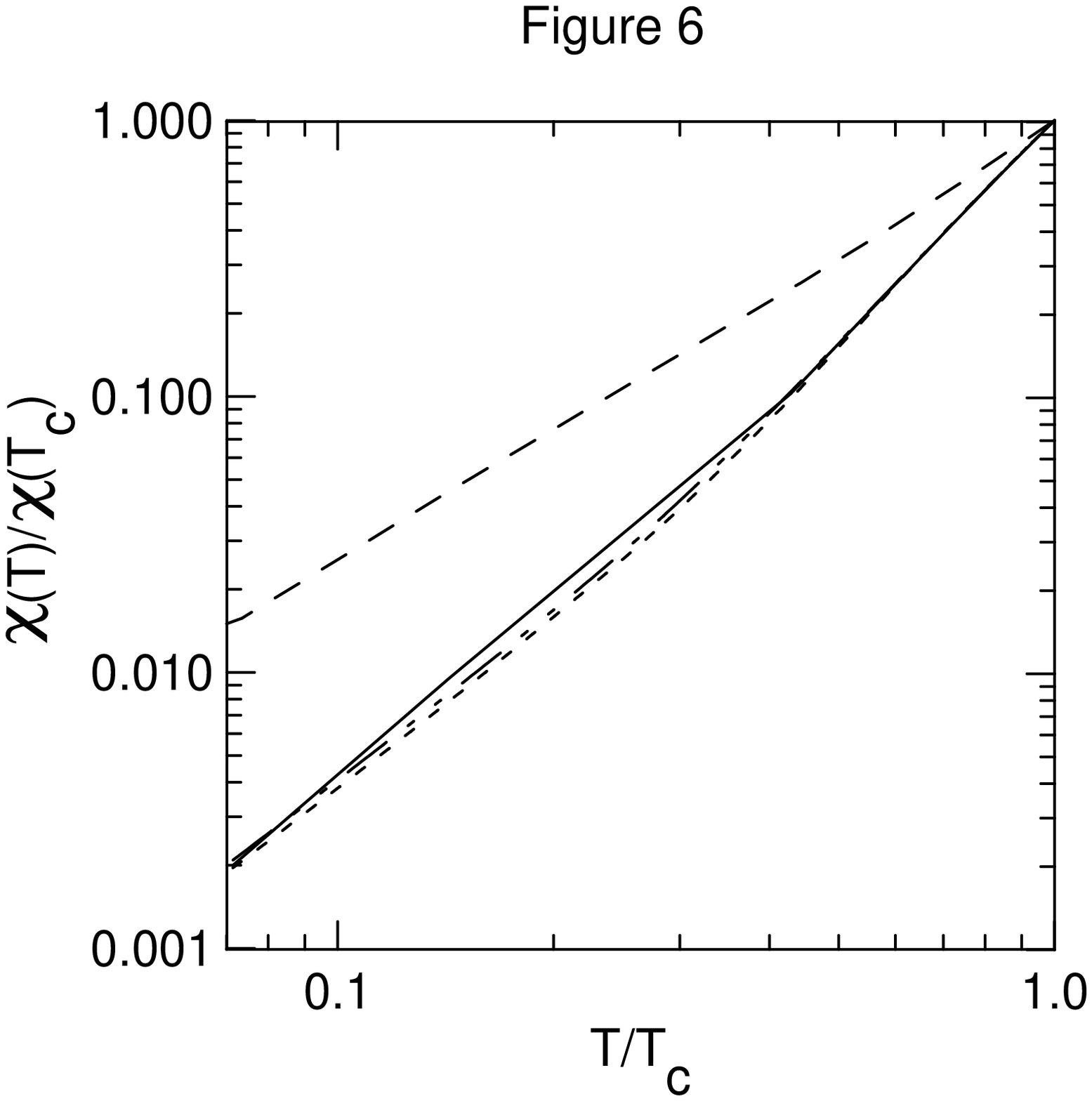}{1.0}    
\vskip -1.5cm

{{\bf Fig. 6.}   Susceptibility  ratio $\chi(T)/\chi(T_c)$ versus $T/T_c$  
in the presence of orthorhombic distortion: 
pure $d_{x^2-y^2}$ wave (solid line),
pure $d_{xy}$ wave (dashed line), model 1(a) (dotted line), and 
model 1(b) (dashed-double-dotted line).}

Next we exhibit  the temperature dependence of spin susceptibility
(defined in Ref. \cite{c})
 in Figs. 5 and 6 where we also plot the results for pure
$d_{x^2-y^2}$ and $d_{xy}$ waves for comparison.  In Figs. 5 and 6, we
show the results for models 1 and 2 on square lattice and with
orthorhombic distortion, respectively.  For pure $d_{x^2-y^2}$ and
$d_{xy}$ waves we obtain power-law dependencies on temperature. The
exponent for this power-law scaling was independent of critical
temperature $T_c$ but varied from a square lattice to that with an
orthorhombic distortion. For $d_{x^2-y^2}$ wave, the exponent for square
lattice (orthorhombic distortion, $\beta$ = 0.95)  is 2.6 (2.4).  For
$d_{xy}$ wave, the exponent for square lattice (orthorhombic distortion,
$\beta$ = 0.95)  is 1.1 (1.6).  For the mixed $d_{x^2-y^2}+d_{xy}$ wave
these exponents are nearly identical to the pure $d_{x^2-y^2}$ wave case.
Hence, by studying the temperature dependency of spin susceptibility, it
will be impossible to detect the phase transition at $T=T_{c1}$ from a
$d_{x^2-y^2}$ wave to a $d_{x^2-y^2}+d_{xy}$ wave, at least within the
present tight-binding model.

In conclusion, we have studied the $ d_{x^2-y^2}+d_{xy}$-wave
superconductivity employing the two-dimensional tight binding BCS model on
square lattice and also on a lattice with orthorhombic distortion and
confirmed a second second-order phase transition at $T=T_{c1}$ in the
presence of a weaker $d_{xy}$ wave.  This phase transition is marked by a
jump in the specific heat at $T=T_{c1}$.  We have kept the $s$- and
$d$-wave couplings in such a domain that a coupled $
d_{x^2-y^2}+d_{xy}$-wave solution is allowed. The
$d_{x^2-y^2}+d_{xy}$-wave state is similar to a $d_{x^2-y^2}$-wave-type
state with nodes on the Fermi surface in the order parameter. 
Consequently, we find power-law temperature dependencies of specific heat
and spin susceptibility in the $d_{x^2-y^2}+d_{xy}$ wave. The exponents of
these power laws for the mixed $d_{x^2-y^2}+d_{xy}$ wave are very close to
those for the pure $d_{x^2-y^2}$ wave.

The work was supported by the CNPq and FAPESP.


\begin{references}

 

\bibitem{n1}H. Ding, Nature {\bf 382}, 51 (1996); D. J. Scalapino, Phys. 
Rep.  {\bf 250}, 329 (1995). 


\bibitem{h} W. Hardy {\it et al.}, Phys. Rev. Lett. {\bf 70}, 3999 (1993);
K. A. Moler {\it et al.}, {\it ibid.} {\bf 73}, 2744 (1994); K. Gofron {\it
et al.}, {\it ibid.} {\bf 73}, 3302 (1994).

\bibitem{t1} M. Prohammer, A. Perez-Gonzalez, and J. P. Carbotte, Phys. Rev. B
{\bf 47}, 15152 (1993);  J. Annett, N. Goldenfeld, and S. R. Renn, {\it
ibid.} {\bf 43} 2778 (1991); N. Momono and M. Ido, Physica C {\bf 264},
311 (1996); M. Houssa and M. Ausloos, {\it ibid.} {\bf 265}, 258 (1996).

\bibitem{c} 
S.  K. Adhikari and A. Ghosh, Phys. Rev. B  {\bf 55}, 1110 (1997); J. Phys.:
Cond. Mat. {\bf 10}, 135 (1998);  A. Ghosh and S. K. Adhikari, Euro. Phys. J
B {\bf 2}, 31 (1998).

\bibitem{6}A. E. Ruckenstein, P. J. Hirschfeld, and J. Apel, Phys. Rev. B 
{\bf 36}, 857 (1987); G. Kotliar, {\it ibid.} {\bf 37}, 3664 (1988). 


\bibitem{5}A. G. Sun, D. A. Gajewski, M. B. Maple, and R. C. Dynes, Phys.
Rev. Lett. {\bf 72}, 2267 (1994); A. G. Sun{\it  et al.}, Phys. Rev. B 
{\bf 52}, R15731 (1995).


\bibitem{8}J. H. Xu, J. L. Shen, J. H. Miller, Jr., 
and C. S. Ting, Phys. Rev. Lett. {\bf 73}, 2492 (1994).


\bibitem{K}K. A. Kouznetsov {\it et al.}, Phys. Rev. Lett. {\bf 79}, 3050
(1997).

\bibitem{S} S. Sridhar {\it et al.} Physica C {\bf 282}, 256 (1997).

\bibitem{7}Jian Ma {\it et al.}, Science {\bf 267}, 83 (1997); 
P. Chaudhari and
S. Y. Lin, Phys.
Rev. Lett. {\bf 72}, 1084 (1994).

\bibitem{bj1}J. Betouras and R. Joynt, Europhys. Lett. {\bf 31}, 119
(1995).


\bibitem{Kr} K. Krishana, N. P. Ong, Q. Li, G. D. Gu, and N. Koshizuka,
Science {\bf 277}, 83 (1997).

\bibitem{L}R. B. Laughlin, Phys. Rev. Lett. {\bf 80}, 5188 (1998); 
M.
Franz and Z. Te\'sanovi\'c, ibid . {\bf 80}, 4763
(1998). 

\bibitem{az} A. Balatsky, Phys. Rev. Lett. {\bf 80}, 1972
(1998).


\bibitem{bj2}J. Betouras and R. Joynt, Phys. Rev. B {\bf 57}, 11752
(1998).

\bibitem{9a} J.-X. Zhu, W. Kim, and C. S. Ting, Phys. Rev. B {\bf 57},
13410 (1998);  M. Liu, D. Y. Xing, and Z. D. Wang, Phys. Rev. B {\bf 55},
3181 (1997); Y. Ren, J. H. Xu, and C. S. Ting, {\it ibid.} {\bf 53}, 2249
(1996);  M. T. Beal-Monod and K. Maki, { Physica C} {\bf 265}, 309 (1996);
 A. Ghosh and S.  K.  Adhikari, ibid.  {\bf 309}, 251 (1998). 


\bibitem{P}P. Monthoux, A. V. Balatsky, and D. Pines, Phys. Rev. B
{\bf 46}, 14803 (1992).

\bibitem{fn} R. Fehrenbacher and M. R. Norman, Phys. Rev. Lett.
{\bf 74}, 3884 (1995).




\end{references}
\end{document}